# Light Field Synthesis by Training Deep Network in the Refocused Image Domain

Chang-Le Liu, Kuang-Tsu Shih, Jiun-Woei Huang, and Homer H. Chen, *Fellow, IEEE*

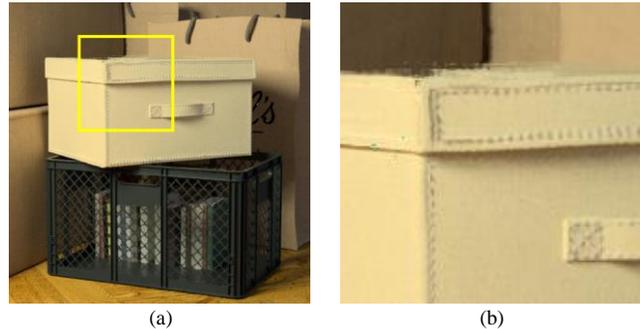

Fig. 1. Illustration of artifacts generated by a network trained to minimize individual view loss. (a) A new view synthesized from a light field in the HCI dataset. (b) The blowup view of a small region.

*Abstract*—Light field imaging, which captures spatial-angular information of light incident on image sensors, enables many interesting applications such as image refocusing and augmented reality. However, due to the limited sensor resolution, a trade-off exists between the spatial and angular resolutions. To increase the angular resolution, view synthesis techniques have been adopted to generate new views from existing views. However, traditional learning-based view synthesis mainly considers the image quality of each view of the light field and neglects the quality of the refocused images. In this paper, we propose a new loss function called refocused image error (RIE) to address the issue. The main idea is that the image quality of the synthesized light field should be optimized in the refocused image domain because it is where the light field is viewed. We analyze the behavior of RIE in the spectral domain and test the performance of our approach against previous approaches on both real (INRIA) and software-rendered (HCI) light field datasets using objective assessment metrics such as MSE, MAE, PSNR, SSIM, and GMSD. Experimental results show that the light field generated by our method results in better refocused images than previous methods.

*Index Terms*—**light field, view synthesis, CNN, image refocusing**

## I. Introduction

LIGHT field imaging enables users to refocus and change view-angle. A light field can be captured by a light field camera equipped with a microlens array [1], [2] or by a camera array [3]. The angular resolution of the light field captured by the former is limited by the number of pixels covered by microlens and the latter by the number of cameras. In practice, only limited angular resolution is available. To increase the angular resolution, view synthesis (interpolation) is often adopted to generate new views from the existing light field.

A view synthesis approach that has been commonly adopted first estimates the disparity map (or depth map) of the light field and then uses it to generate new views in between the existing views by warping the existing light field [4]–[7]. Recently, deep learning has been applied to view synthesis of light field [8]–[10]. It involves several convolutional neural networks (CNN) to estimate the disparity map and refine the warped new views using an end-to-end training strategy. The output is a light field denser than the input. The loss functions used are mostly $L1$ or $L2$ metric. Furthermore, these metrics for evaluating the quality of the synthesized light field is performed view-wise (that is, view by view) in the 4D light field domain.

In our view, the quality of view synthesis for light field should be evaluated in the refocused image domain, because what matters ultimately is the perceived image, not the volume of 4D raw data. Since the perceived image is a refocused image, we believe a light field synthesis optimized in the refocused domain would generate better refocused images. Fig. 1 illustrates a common drawback of conventional light field synthesis. View-wise optimization usually results in artifact at the region where occlusion occurs.

To take the refocused image quality into consideration, we propose to add refocused image error (RIE) to the traditional view-wise loss function as a regularization term. RIE encourages the network to focus on the light field quality in the refocused image domain. It results in a dense light field, from which a high-quality refocused image can be generated.

The contributions of this paper can be summarized as follows:
- To our best knowledge, this is the first work that considers refocused image quality for light field synthesis and optimizes the deep network in the refocused image domain.
- We analyze the proposed refocused image error in the spectral domain and show the relation between evaluating $L2$ loss in

This work was supported in part by a grant from the Ministry of Science and Technology of Taiwan under Contract 106-2221-E-002-201-MY3, and in part by grants from National Taiwan University under Contracts CC-NTU: 108L891808 and 108L880502.

C.-L. Liu is with the Graduate Institute of Communication Engineering, National Taiwan University, Taipei 10617, Taiwan (e-mail: b05901017@ntu.edu.tw).

K.-T. Shih is with the Graduate Institute of Communication Engineering, National Taiwan University, Taipei 10617, Taiwan (e-mail: shihkt@gmail.com)

J.-W. Huang is with the Institute of Applied Mechanics, National Taiwan University, Taipei 10617, Taiwan (jwhuang@ms2.hinet.net)

H. H. Chen is with the Department of Electrical Engineering, Graduate Institute of Communication Engineering, and Graduate Institute of Networking and Multimedia, National Taiwan University, Taipei 10617, Taiwan (e-mail: homer@ntu.edu.tw).

4D light field domain and that in the refocused image domain.
- We demonstrate that taking refocused image quality into consideration improves the performance of deep learning-based light field synthesis.

The rest of the paper is organized as follows. We review the related work in Sec. II and introduce the notation used in this paper in Sec. III. The proposed regularization and related analysis are described in Sec. IV. The experiment setting and the results are described in Sec. V. Finally, the concluding remarks and future work discussion are provided in Sec. VI.

## II. RELATED WORK

### A. Light Field Imaging

A light field [11] records the spatial-angular information of the light rays coming from different view angles. It contains angular information unavailable in the traditional 2D image data. Using the spatial-angular information, we may perform image refocusing to make any object in the scene in focus [1]. A light field can be captured by using an array of cameras [3] or a camera with a lenslet array [1]. The latter allows miniaturization of the device for consumer electronics [12], [13]. However, both types of light field cameras have limited angular resolution. The angular resolution of the former is limited by the number of cameras and the latter by the number of corresponding pixels of a lenslet.

### B. Image Quality Assessment

The loss function used in the training process usually involves image quality assessment and decides how a neural network is to be optimized. Hence, an appropriate loss function helps generate high-quality output images. Although the mean square error (MSE) is one of the most widely used and simplest image quality metrics, it is not necessarily consistent with human evaluation. Many perceptual quality metrics have been developed [14]. Especially, perceptual losses based on high-level features extracted from some pre-trained networks have been developed [15]–[19]. For example, Johnson *et al.* [15] and Bruna *et al.* [18] proposed to use high-level features extracted from VGG network [20] to evaluate the quality of output images. Note that these 2-D image quality assessment methods are not designed for light field.

Meng *et al.* [17] adopted the concept of perceptual loss and proposed to extract the VGG features of each view of a light field and compute the feature distance (difference) as the loss function for training. However, the focus of this method is still on the optimization of the light field view-by-view without considering the perceived quality.

### C. View Synthesis for Light Field

View synthesis [3] has been developed to increase the angular resolution of a light field. Generally, view synthesis methods for the light field could be classified into two types.

The first type of methods [4], [5], [7] first estimate the depth information and then warp the existing views to generate new views by multi-view stereo algorithms [21], [22]. It is a depth-dependent process. The convolutional neural network (CNN) has also been adopted for view synthesis. Kalantari *et al.* proposed to use CNN to evaluate disparity information from the four corner views in an input light field. Another CNN was used to refine the new views generated the existing views and to estimate the depth map [9]. Srinivasan *et al.* extended this method to synthesize a light field only from the central view in the light field using an extra constraint on the consistency of ray depths [8].

TABLE I
NOMENCLATURE

| Symbol | Description |
|---|---|
| $L(x, y, s, t)$ | 4-D light field. The pairs $(x, y)$ and $(s, t)$ represent the spatial and angular coordinates, respectively. |
| $L_\mathbf{s}(\mathbf{x})$ | Sub-aperture image at $\mathbf{s} = (s, t)$. That is, the view captured at the angular coordinates $(s, t)$, where $\mathbf{x} = (x, y)$. |
| $\mathfrak{F}$ | Non-unitary Fourier transform. We drop the constant coefficient for simplicity. |
| $\mathfrak{F}^{-1}$ | Inverse Fourier transform. We drop the constant coefficient for simplicity. |
| sinc($x$) | Unnormalized sinc function sin($x$)/$x$ |
| $G_\theta$ | Neural network parameterized by $\theta$ |
| $g(r)$ | Gaussian function exp($-r^2$) |

The second type of method synthesizes a new image without depth information. This is made possible by limiting the configuration of input views to some specific patterns [23], [24]. Although the depth information is not required, the constraint on the configuration of input views limits the application of such methods. To overcome this problem and improve the performance for occlusion regions and non-Lambertian surfaces (which do not reflect light equally in all directions), for which depth-dependent methods often fail, Wang *et al.* proposed Pseudo 4DCNN [10] that adopts 3D CNN [25] to extract the 3D volume features by alternatively fixing an angular dimension of the input light field. That is, the Pseudo 4DCNN directly generates a dense light field by upsampling the input light field using deconvolution.

Although the above methods solve the light field view synthesis problem to a certain extent, the quality of refocused images is not explored. In this paper, we propose to add a regularization term called refocused image error to the loss function. It encourages the network to generate high-quality refocused images.

## III. NOTATION

We consider the 4D light field proposed by Levoy and Hanrahan [3] and define the symbols used in the paper in Table I. Note that $s$ and $t$ are finite in the most cases in practice; therefore, we assume each of them is bounded by $N$ without loss of generality.

We also introduce the *shift-and-add* operation, which is a basic operation for image refocusing. In this operation, we first shift each sub-aperture image $L_\mathbf{s}(\mathbf{x})$ by $\Delta \mathbf{x} = r\mathbf{s}$ and then average all shifted image to generate a refocused image $R$. More specifically,

$$R(L, r)(\mathbf{x}) = \frac{1}{(2N+1)^2} \sum_{\mathbf{s}=(-N,-N)}^{(N,N)} L_\mathbf{s}(\mathbf{x} + r\mathbf{s}). \quad (1)$$

Note that a larger $r$ means refocusing farther from the camera and the sign of $r$ decides the position of the refocused focal point with respect to the original focal point: positive means refocusing at a farther object and negative at a closer object.

## IV. REFOCUSED IMAGE ERROR

In this section, we first introduce the refocused image error and provide related analysis in the frequency domain.

### A. Refocused Image Error

Assume there exists a desired light field $L$ and a set of images $S$ sampled from $L$. Given input $S$, we want to train a neural network $G$ parameterized by $\theta$ to predict a light field $G_\theta(S)$ that is as similar as possible to $L$. Note that the proposed method has no restriction on the type of input, even though the input of the network is a set of views of the desired light field in this work. The only limitation is that the output must be a light field. Mathematically, the network $G$ is trained to minimize the loss between $G_\theta(S)$ and $L$ as follows:

$$\theta = \arg\min_\theta \mathcal{L}(G_\theta(S), L) \qquad (2)$$

Traditionally, the loss function $\mathcal{L}$ is chosen to be the mean-squared error (MSE) or the mean absolute error (MAE) between every image in $L$ and $G_\theta(S)$. These view-wise errors (VWE) can be defined as follows:

$$\text{VWE}_1(G_\theta(S), L) = \sum_{\mathbf{s}=(-N,-N)}^{(N,N)} \text{MAE}(G_\theta(S)_\mathbf{s}, L_\mathbf{s}) \qquad (3)$$

$$\text{VWE}_2(G_\theta(S), L) = \sum_{\mathbf{s}=(-N,-N)}^{(N,N)} \text{MSE}(G_\theta(S)_\mathbf{s}, L_\mathbf{s}) \qquad (4)$$

This kind of loss functions only encourages the network to perform well on each sub-aperture image without considering the quality of refocused images generated from the predicted light field $G_\theta(S)$.

Instead, we propose the unweighted continuous refocused image error (UCRIE),

$$\text{UCRIE}_1(G_\theta(S), L) \\ = \frac{1}{2D} \int_{-D}^{D} \text{MAE}(R(G_\theta(S), r), R(L, r)) dr, \qquad (5)$$

$$\text{UCRIE}_2(G_\theta(S), L) \\ = \frac{1}{2D} \int_{-D}^{D} \text{MSE}(R(G_\theta(S), r), R(L, r)) dr, \qquad (6)$$

where $D$ denotes the maximal value of $r$ in the *shift-and-add* operation. Intuitively, UCRIE$_2$ and UCRIE$_1$ correspond to MSE and MAE, respectively.

### B. Spectral Domain Analysis

Here we analyze the characteristics of UCRIE in the spectral domain. First, we may rewrite UCRIE$_2$ using Fourier transform and Plancherel's formula as follows:

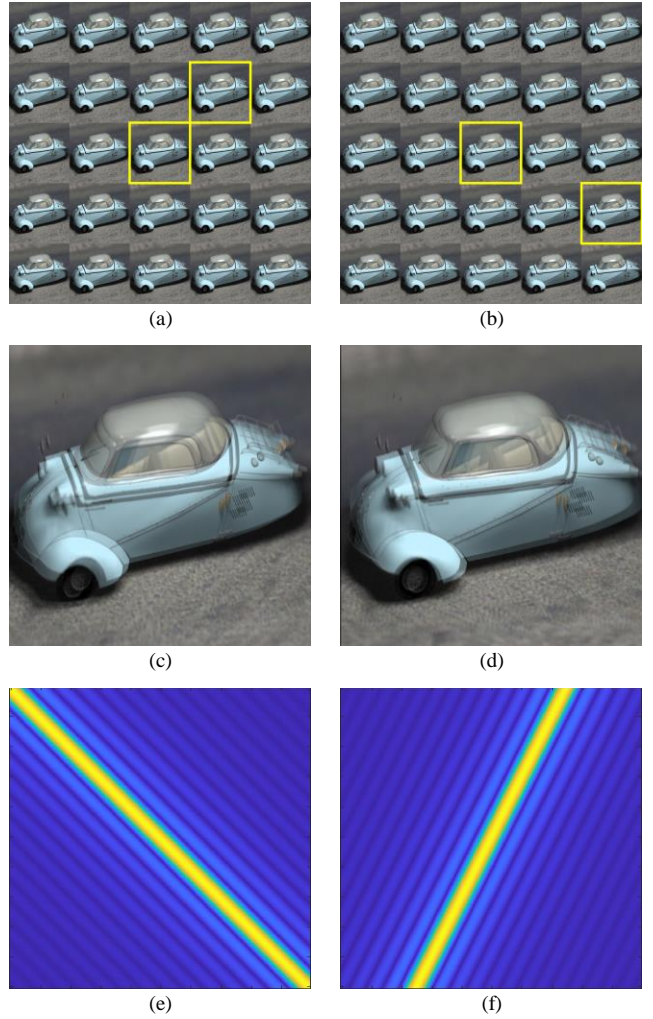

Fig. 2. (a) and (b) Different sub-aperture images chosen from a 5×5 light field. (c) and (d) The superposition of two sub-aperture images in (a) and (b). (e) A 135-degree orientated sinc function. (f) A 60-degree orientated sinc function.

$$\text{UCRIE}_2(G_\theta(S), L) = \\ \frac{1}{(2N+1)^4} \sum_{\mathbf{s},\mathbf{t}} \sum_\omega \mathcal{E}_\mathbf{s}(\omega)\mathcal{E}_\mathbf{t}(\omega)\text{sinc}(D\omega^T(\mathbf{s}+\mathbf{t})), \qquad (7)$$

where $\mathcal{E}_\mathbf{s} = \mathfrak{F}\{G_\theta(S)_\mathbf{s} - L_\mathbf{s}\}$ denotes the spectrum of the error of the sub-aperture image at $\mathbf{s}$. The derivation of Eq. (7) is given in Appendix. Eq. (7) suggests that the unweighted refocused continuous refocused image error measures the error filtered by a low-pass sinc filter.

By the definition of MSE, we can also rewrite (4) in frequency domain as

$$\sum_{\mathbf{s}=(-N,-N)}^{(N,N)} \text{MSE}(L_\mathbf{s}, G_\theta(S)_\mathbf{s}) = \sum_{\forall \mathbf{s},\mathbf{t},\mathbf{s}=\mathbf{t}} \sum_\omega \mathcal{E}_\mathbf{s}(\omega)\mathcal{E}_\mathbf{t}(\omega). \qquad (8)$$

We can see that view-wise MSE is a simplification of Eq. (11). Compare UCRIE$_2$ with traditional view-wise MSE, we can see there are two main differences. First, the traditional view-wise MSE does not consider the correlation between two different views captured from different view angles. Second, the error measured by UCRIE$_2$ is filtered by a directional filter. That is, the weighting for each $\sum_{\omega} \mathcal{E}_s(\omega)\mathcal{E}_t(\omega)$ in Eqs. (7) depends on the vector value of **s** + **t**. For example, let **s** = (0, 0) and assume **s** + **t** = (1, 1), without loss of generality. Then $L_t$ is in the top right of $L_s$, as marked in Fig. 2(a). According to Eq. (7), we see the weight of $\mathcal{E}_s(\omega)\mathcal{E}_t(\omega)$ is a 135-degree orientated low-pass sinc filter, as shown in Fig. 2(e). This is reasonable. If we *shift-and-add $L_s$ and $L_t$*: $L_t$ moves toward (away) from $L_s$ when $r$ is positive (negative) in the 45-degree orientation, we get an image with motion blur in the 45-degree orientation, as shown in Fig. 2(c). In another case, assume **s** + **t** = (2, −1), as shown in Fig. 2(b). Then the weight is a 60-degree orientated sinc filter, as shown in Fig. 2(f), and we get an image with motion blur in the other orientation, as shown in Fig. 2(c).

Because the sinc filter exhibits undesirable oscillatory behavior, which can also be observed in Fig. 2(e) and 2(f), we weight UCRIE by a Gaussian function $g(r)$ and define the continuous refocused image error (CRIE) as follows:

$$\text{CRIE}_1(G_\theta(S), L) = \frac{1}{2D} \int_{-\infty}^{\infty} g(r)\text{MAE}(R(G_\theta(S), r), R(L, r))dr, \quad (9)$$

$$\text{CRIE}_2(G_\theta(S), L) = \frac{1}{2D} \int_{-\infty}^{\infty} g(r)\text{MSE}(R(G_\theta(S), r), R(L, r))dr. \quad (10)$$

Note that the definite integral is replaced with improper integral in CRIE for simplification in the spectral domain. (By the definition of $g(r)$, this makes a trivial difference when $D$ is larger than 3.) Incidentally, we adopt definite integral for UCRIE to prevent it from diverges when $D$ goes to infinity. We can, as what we do for Eqs (6) and (7), rewrite (10) in frequency domain:

$$\text{CRIE}_2(G_\theta(S), L) = \frac{1}{(2N+1)^4} \sum_{s,t} \sum_{\omega} \mathcal{E}_s(\omega)\mathcal{E}_t(\omega) \frac{\sqrt{\pi}}{2D} e^{-0.25(\omega^T(s+t))^2} \quad (11)$$

By this way, the sinc filter (in UCRIE$_2$) is replaced with a Gaussian filter, which is non-negative and non-oscillatory in the spectral domain.

For CRIE$_1$ and UCRIE$_1$, we can analyze them in a similar way. By using Chebyshev approximation [26] and omitting terms higher than the fourth order, we have the following:

$$\text{CRIE}_1(G_\theta(S), L) = \frac{-1}{\pi} \frac{1}{2D} (\frac{1}{2N+1})^2$$
$$\int_{-\infty}^{\infty} e^{-r^2} \left\{ \sum_x \left( \Sigma_{k=1}^{\infty} \frac{(-1)^k T_{2k}(\mathfrak{r}(x,r))}{-1+4k^2} \right) \right\} dr, \quad (12)$$
$$\mathfrak{r}(x,r) = R(G_\theta(S), r) - R(L, r),$$

where $T_{2k}$ is a Chebyshev polynomial of the first kind. By expanding (12), we can find that CRIE$_1$ includes CRIE$_2$.

Although $r$ in UCRIE and CRIE can have infinitely many values, evaluating an equation on infinitely many points is difficult. In practice, the following losses, called refocused image errors (RIEs) in this work, are more appropriate for deep learning and quality evaluation tasks:

$$\text{RIE}_1(G_\theta(S), L) = \frac{1}{2D} \sum_{r=-D/s}^{D/s} g(r)\text{MAE}(R(G_\theta(S), sr), R(L, sr)), \quad (13)$$

$$\text{RIE}_2(G_\theta(S), L) = \frac{1}{2D} \sum_{r=-D/s}^{D/s} g(r)\text{MSE}(R(G_\theta(S), sr), R(L, sr)), \quad (14)$$

where $s$ is the step interval of the summation.

## V. EXPERIMENTAL SETUP AND RESULTS

In this section, we first describe the network architecture used in our experiment for view synthesis and the experimental setup. Then, we describe the results of our experiments for testing the performance of the proposed RIE for light field synthesis.

### A. Network Architecture

We consider a neural network trained through backpropagation for light field synthesis. Specifically, the deep neural network architecture proposed by Srinivasan1 *et al.* [8] was used in the experiments. Since synthesizing a light field from multiple views is more robust than from a single view, we fed a total of five views into the network as inputs. This architecture consists of two sub-networks, both fully convolutional networks. The first sub-network estimated the depth map of each new view, based on which an approximate Lambertian light field was synthesized by warping the central view. The second sub-network predicted a residual light field to be added to the synthesized Lambertian light field and handled occluded parts and non-Lambertian effects [8]. In our experiment, we trained two networks to minimize the following two loss functions, one for each network:

$$\text{VWE}_1(G_\theta(S), L) + \lambda_R \text{RIE}_1(G_\theta(S), L), \quad (15)$$
$$\text{VWE}_2(G_\theta(S), L) + \lambda_R \text{RIE}_2(G_\theta(S), L), \quad (16)$$

where $\lambda_R$ denotes the RIE regularization parameters. The training scheme is shown in Fig. 3. For comparison, we trained another two networks to minimize Eqs. (3) and (4).

Note that at first glance, using Eqs. (7) and (11) as the loss function may seem appropriate for evaluating Eqs. (7) and (8) on infinitely many samples of $r$. However, using them for training usually leads to unstable results. This is why Eqs. (13) and (14) are used in Eqs. (15) and (16), respectively, for network training in our experiments.

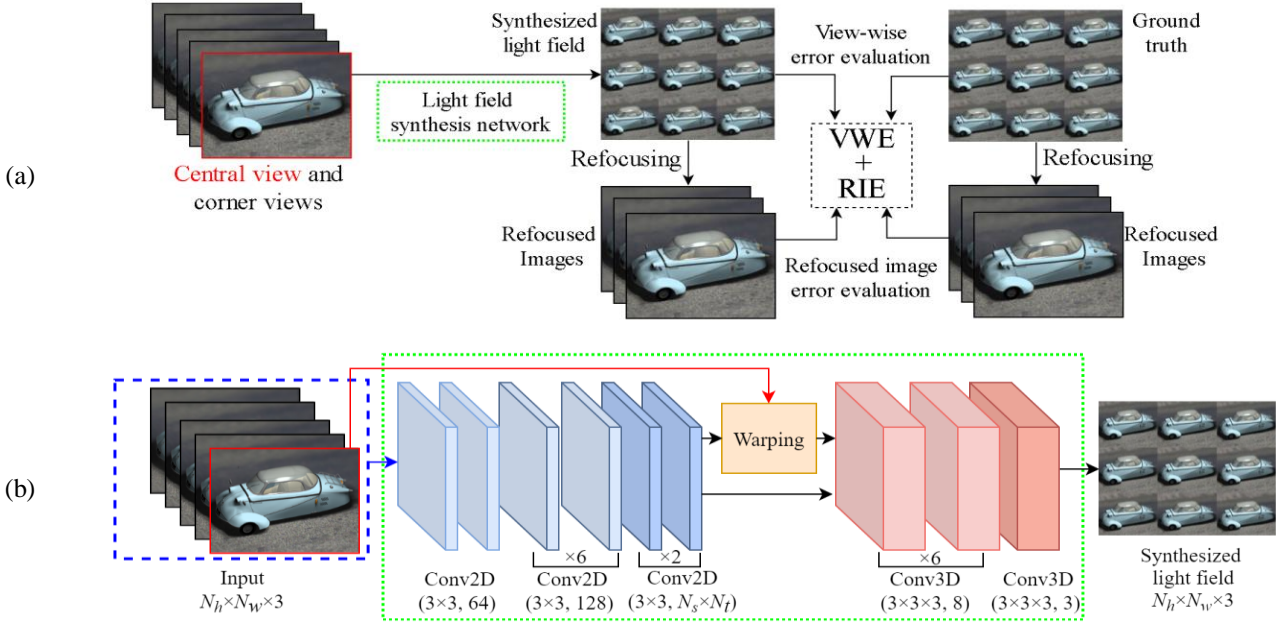

Fig. 3. (a) The training procedure for the proposed refocused image error between the synthesized light field and the ground-truth. In addition to the view-wise error, the light field synthesis network is trained end-to-end to minimize the refocused image error. (b) Architecture of the light field synthesis network. 2-D convolutional and 3-D convolutional layers are represented by bluish and reddish rectangular cuboids, respectively, and the parameters are denoted as Conv2D ($p_1$, $p_2$, $p_c$) and Conv3D ($p_1$, $p_2$, $p_3$, $p_c$), where $p_1$, $p_2$ and $p_3$ denote the kernel size, $p_c$ denotes the number of filters channels, and ($N_h$, $N_w$) and ($N_s$, $N_t$), respectively, denote the image size of the input and angular resolution of the output.

## B. Experimental Setup

We evaluated the network performance in terms of following standard objective metrics: MAE, MSE, the peak signal-to-noise ratio (PSNR), the gray-scale structural similarity (SSIM) [27], and the gradient magnitude similarity deviation (GMSD) [28]. SSIM is one of perception-based image quality metric that considers image luminance, contrast, and structure similarity. The definition of SSIM is described below:

$$\text{SSIM}(x, y) = [l(x, y)]^\alpha [c(x, y)]^\beta [s(x, y)]^\gamma$$
$$l(x, y) = \frac{2\mu_x \mu_y + C_1}{\mu_x^2 + \mu_y^2 + C_1}$$
$$c(x, y) = \frac{2\sigma_x \sigma_y + C_2}{\sigma_x^2 + \sigma_y^2 + C_2} \qquad (17)$$
$$s(x, y) = \frac{\sigma_{xy} + C_3}{\sigma_x \sigma_y + C_3}$$

where $\mu_x$, $\mu_y$, $\sigma_x$, $\sigma_y$, and $\sigma_{xy}$ are the local means, standard deviations, and cross-covariance for distorted images $x$ and the reference image $y$. $C_1$, $C_2$ and $C_3$ are variables to stabilize the division. GMSD is another image quality metric that uses the standard deviation of the pixel-wise gradient magnitude similarity (GMS) to evaluate image quality,

$$\text{GMS}(\mathbf{x}) = \frac{2m_r(\mathbf{x})m_d(\mathbf{x}) + c}{m_r^2(\mathbf{x}) + m_d^2(\mathbf{x}) + c},$$
$$\text{GMSD} = \sqrt{\frac{1}{M}\sum_\mathbf{x}(\text{GMS}(\mathbf{x}) - \frac{1}{M}\sum_\mathbf{x}\text{GMS}(\mathbf{x}))^2}, \qquad (18)$$

where $M$ is the number of pixels, $m_r$ and $m_d$ are gradient magnitudes of the reference and the distorted images, respectively. The worse the distorted image, the higher GMSD is. We used SSIM and GMSD because these image quality metrics have a high agreement with the subjective experimental results [29] for light field quality evaluation.

We trained the networks on two light field datasets: the synthetic light fields (HCI dataset) [30] and the real light fields (INRIA dataset [31]). We partitioned the HCI dataset (24 light fields in total) into 16 light fields for training and 8 for testing. Likewise, we partitioned the INRIA dataset (59 light fields in total) into 43 light fields for training and 16 for testing. Both HCI and INRIA datasets have a spatial resolution of $512 \times 512$. But the former has an angular resolution of $9 \times 9$, and the latter $7 \times 7$. Furthermore, we extracted sub-lightfields of $5 \times 5$ views from each light field in the datasets, and hence a total of $(9 - (5 - 1))^2 = 25$ sub-lightfields were extracted from the HCI dataset and $(7 - (5 - 1))^2 = 9$ from the INRIA dataset. For each sub-light field, we used the central view and the four corner views as input to the view synthesis network. The output was a $3 \times 3$ light field, as shown in Fig. 3.

For each new view, the corresponding view at the same position in the original light field was used as the ground-truth. For the RIE evaluation, we set $D = 2.5$, $s = 0.25$ and $\lambda_R = 1$. All neural network models (with 2.5M trainable parameters) were trained from scratch using the Adam optimization algorithm [32] with default parameter values $\beta_1 = 0.9$, $\beta_2 = 0.999$, $\epsilon = 1e{-}08$. The learning rates and the numbers of epochs of all models are 0.001 and 300, respectively.

Table II
4D LIGHT FIELD QUALITY COMPARISON OF OUR APPROACH WITH TRADITIONAL LOSS FUNCTION USING HCI DATASET

| Quality Metrics | Loss Functions | | | |
|---|---|---|---|---|
| | $VWE_2$ | $VWE_2+RIE_2$ | $VWE_1$ | $VWE_1+RIE_1$ |
| MAE | 0.0180 | **0.0176** | 0.0152 | **0.0156** |
| MSE | 0.0032 | **0.0029** | 0.0012 | **0.0010** |
| PSNR | 27.310 | **28.098** | 30.340 | **30.727** |

Table III
4D LIGHT FIELD QUALITY COMPARISON OF OUR APPROACH WITH TRADITIONAL LOSS FUNCTION USING INRIA DATASET

| Quality Metrics | Loss Functions | | | |
|---|---|---|---|---|
| | $VWE_2$ | $VWE_2+RIE_2$ | $VWE_1$ | $VWE_1+RIE_1$ |
| MAE | 0.0176 | **0.0094** | 0.0091 | **0.0091** |
| MSE | 0.0030 | **0.0003** | 0.0005 | **0.0004** |
| PSNR | 27.805 | **36.397** | 33.354 | **35.213** |

Furthermore, because the sizes of the HCI and the INRIA datasets are not large, we conducted a similar experiment with k-fold (k = 5) cross-validation on all 24 and 59 light fields of the HCI and the INRIA datasets to verify the effectiveness of RIE. The setting of the experiment is the same as that mentioned above, except that the spatial resolution is down-sampled to 128 × 128 to saving the training time.

### C. Experimental Results

We trained two networks using the proposed loss functions. The one using Eq. (15) as the loss function is referred to as $VWE_1+RIE_1$, and the one using Eq. (16) as the loss function is referred to as $VWE_2+RIE_2$. In addition, two other networks called $VWE_1$ and $VWE_2$ for short were created using Eqs. (3) and (4), respectively, as the loss functions. We also trained another network using the state-of-the-art loss function proposed by *Srinivasan1* et al. [8]. The latter three networks served as the baseline for comparison, and all five networks were tested on the HCI and INRIA datasets. We measured the mean scores of all sub-aperture views using MAE, MSE, and PSNR as the quality metrics and summarized the results in Table II and Table III. We can see the PSNR scores of $VWE_2+RIE_2$ and $VWE_1+RIE_1$ are higher than those of $VWE_2$ and $VWE_1$, respectively. But the MSE scores of the former two are lower than those of the latter two. The MAE score of $VWE_2+RIE_2$ is also lower than that of $VWE_2$. In addition, both $VWE_1+RIE_1$ and $VWE_1$ perform better than $VWE_2+RIE_2$ and $VWE_2$.

Blowup views of the outputs of the trained networks and the corresponding ground-truth are shown in Fig. 5 for the HCI dataset and Fig. 6 for the INRIA dataset. We can see that the results of $VWE_1+RIE_1$ are less noisy and more stable than those of $VWE_1$ and $VWE_2$. From Fig. 5(b), we note that the artifact of $VWE_1$ and $VWE_2$ in the occlusion region is quite pronounced, while the networks $VWE_1+RIE_1$ with the proposed loss function is almost artifact-free. From Fig. 6, we can see that $VWE_1$ and $VWE_2$ generated more noises than $VWE_1+RIE_1$.

In addition to sub-aperture view quality, we evaluated the refocused image quality. The MAE, MSE, PSNR, SSIM, and

Table IV
REFOCUSED IMAGE QUALITY COMPARISON OF OUR APPROACH WITH TRADITIONAL LOSS FUNCTION USING HCI DATASET

| Quality Metrics | Loss Functions | | | |
|---|---|---|---|---|
| | $VWE_2$ | $VWE_2+RIE_2$ | $VWE_1$ | $VWE_1+RIE_1$ |
| MAE | 0.0096 | **0.0091** | 0.0071 | **0.0070** |
| MSE | 0.0008 | **0.0007** | 0.0002 | 0.0002 |
| PSNR | 31.087 | **31.283** | 37.183 | **38.308** |
| SSIM | 0.8243 | **0.8358** | 0.8468 | 0.8415 |
| GMSD | 0.0461 | **0.0371** | 0.0269 | **0.0240** |

Table V
REFOCUSED IMAGE QUALITY COMPARISON OF OUR APPROACH WITH TRADITIONAL LOSS FUNCTION USING INRIA DATASET

| Quality Assessment | Used loss function in training | | | |
|---|---|---|---|---|
| | $VWE_2$ | $VWE_2+RIE_2$ | $VWE_1$ | $VWE_1+RIE_1$ |
| MAE | 0.0093 | **0.0043** | 0.0043 | **0.0041** |
| MSE | 7.08e-4 | **4.00e-5** | 6.82e-5 | **4.12e-5** |
| PSNR | 31.523 | **44.0113** | 41.7215 | **43.8753** |
| SSIM | 0.8271 | **0.9129** | 0.9143 | **0.9163** |
| GMSD | 0.0428 | **0.0055** | 0.0132 | **0.0074** |

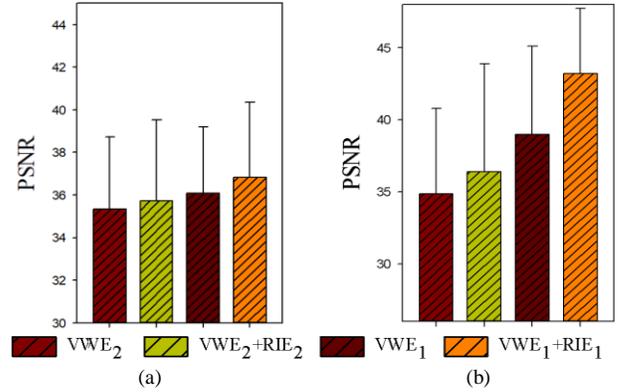

Fig. 4. The PSNR scores of the refocused images in the k-fold cross-validation experiments on (a) HCI and (b) INRIA datasets.

GMSD scores of the refocused images generated by the five networks with r = 0 are shown in Table IV and Table V. We can see from Table IV that $VWE_2+RIE_2$ and $VWE_1+RIE_1$ outperform $VWE_2$ and $VWE_1$ in terms of MAE, PSNR, and GMSD when evaluated on the HCI dataset. Furthermore, the performance of $VWE_2+RIE_2$ and $VWE_1+RIE_1$ is superior to the other three networks by all five metrics when tested on the INRIA dataset. It is worth noting that the average MAE and MSE of the refocused image are lower than the sub-aperture views, and the PSNR scores of the former are higher than the latter. This can be explained by Eq (1) and the triangle inequality. When we obtain a refocused image from a light field, the errors of different sub-aperture views may cancel each other due to the *shift-and-sum* operation.

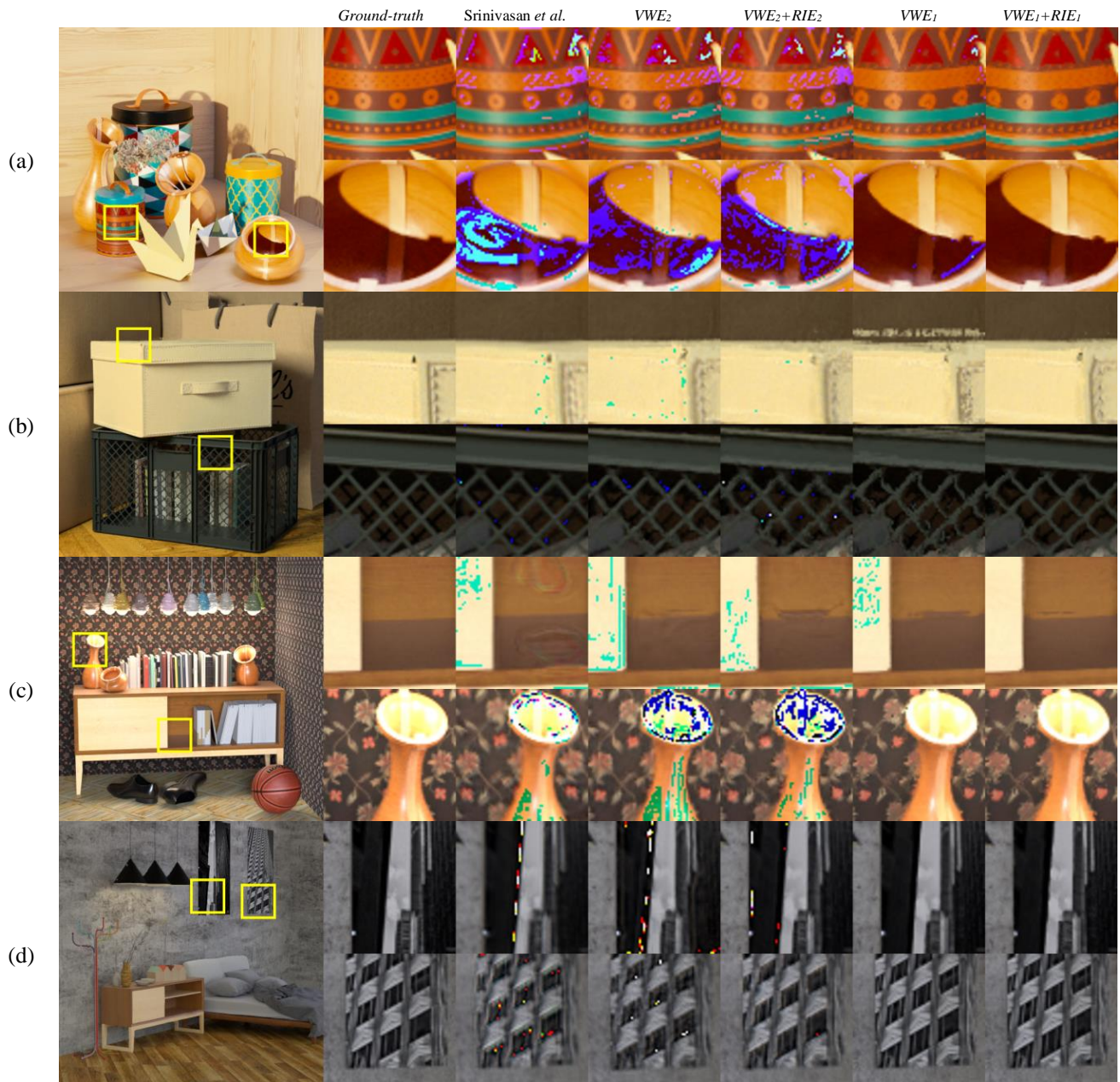

Fig. 5. (a)-(d) Ground-truth and close-up views of experimental results using light fields from the HCI dataset.

The quality of refocused images generated from the synthesized light fields with $r$ ranging from $-2.5$ to $2.5$ is also evaluated. Blowup views of the refocused images generated from the outputs are shown in Fig. 7, and the numeric results are summarized in Fig. 8. Both show that $VWE_2+RIE_2$ and $VWE_1+RIE_1$ perform better than $VWE_2$ and $VWE_1$ by all five image metrics. This is reasonable because a refocused image of a light field can be seen as a linear combination of the shifted sub-aperture images. Therefore, when the sub-aperture images are noisy, the refocused images are likely to be noisy as well. From Figs. 5 − 8 and Tables II − V, we can see that the networks trained by the proposed $VWE_1+RIE_1$ functions indeed give rise to superior light field synthesis in both the 4-D light field domain and the refocused image domain.

The numerical result of k-fold (k = 5) cross-validation is shown in Fig. 4. We can see that $VWE_2+RIE_2$ performs better than $VWE_2$ and that $VWE_1+RIE_1$ performs better than $VWE_1$. The result agrees with that of the previous experiment.

VI. CONCLUSION

In this paper, we have described a novel loss called refocused image error for light field synthesis. It drives a deep network to minimize light field loss in the 4D light field domain and the refocused image domain at the same time, resulting in high-quality refocused images. The superior performance of the proposed loss is supported by a theoretical analysis that shows the refocused image error is related to the summation of the inner products of spectra errors between all view pairs of a synthesized light field. In effect, our technique takes whole light field into consideration.

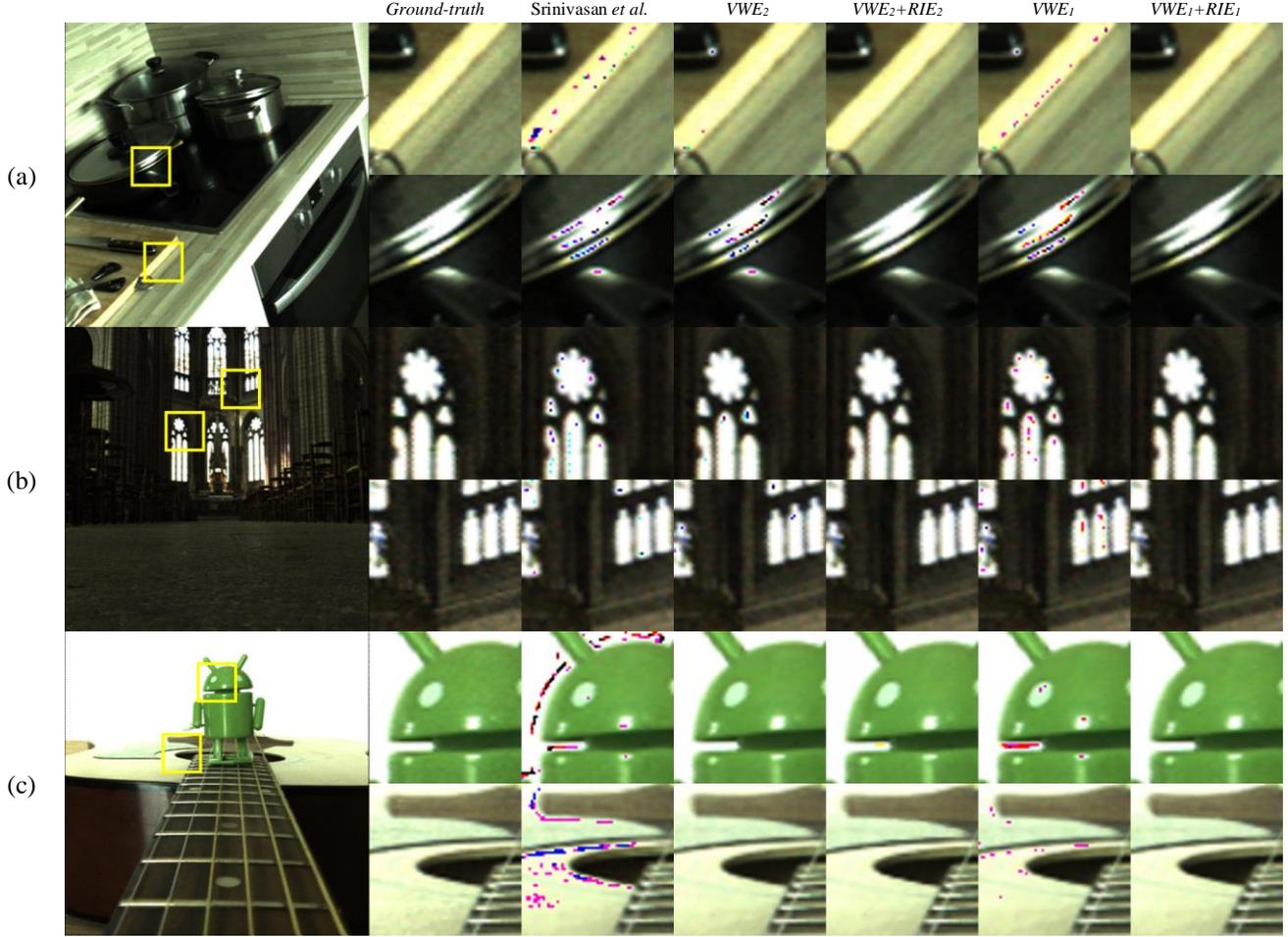

Fig. 6. (a)-(c) Ground-truth and close-up views of experimental results using light fields from the INRIA dataset.

The experimental results using INRIA (a real dataset) and HCI (a software-rendered dataset) clearly show that the proposed regularization is more effective than the conventional one that only considers the individual view quality of a light field. The proposed loss is potentially useful for other light-field related tasks such as light field compression [34] and super-resolution [35]. These topics are worth further investigation in the future.

APPENDIX

For simplicity, let $\hat{L}$ denote the alias of $G_\theta(S)$ and $L_s^{\mathbf{h}} = L_s(\mathbf{x} + \mathbf{h})$. The definitions of UCRIE$_2$ and *shift-and-add* operator in Eqs. (6) and (1) establish the equation:

$$\mathrm{UCRIE}_2(\hat{L}, L)$$
$$= \frac{1}{2D}\int_{-D}^{D} \mathrm{MSE}(R(\hat{L}, r), R(L, r))dr$$
$$= \frac{1}{2D}\int_{-D}^{D} \sum_{\mathbf{x}} (R(\hat{L}, r)(\mathbf{x}) - R(L, r)(\mathbf{x}))^2 dr$$
$$= \frac{1}{2D}\int_{-D}^{D} \sum_{\mathbf{x}} (\frac{1}{(2N+1)^2}\sum_{\mathbf{s}}\hat{L}_{\mathbf{s}}(\mathbf{x} + r\mathbf{s}) - \frac{1}{(2N+1)^2}\sum_{\mathbf{s}}L_{\mathbf{s}}(\mathbf{x} + r\mathbf{s}))^2 dr$$
$$= \frac{1}{2D(2N+1)^4}\int_{-D}^{D} \sum_{\mathbf{x}} (\sum_{\mathbf{s}}(\hat{L}_{\mathbf{s}}^{r\mathbf{s}}(\mathbf{x}) - L_{\mathbf{s}}^{r\mathbf{s}}(\mathbf{x}))^2 dr$$

Because the light fields are finite-valued, we can interchange the order of summation:

$$\frac{1}{2D(2N+1)^4}\int_{-D}^{D} \sum_{\mathbf{x}} (\sum_{\mathbf{s}}(\hat{L}_{\mathbf{s}}^{r\mathbf{s}}(\mathbf{x}) - L_{\mathbf{s}}^{r\mathbf{s}}(\mathbf{x}))^2 dr$$
$$= \frac{1}{2D(2N+1)^4}\int_{-D}^{D} \sum_{\mathbf{s},\mathbf{t}}\sum_{\mathbf{x}} (\hat{L}_{\mathbf{s}}^{r\mathbf{s}}(\mathbf{x}) - L_{\mathbf{s}}^{r\mathbf{s}}(\mathbf{x}))(\hat{L}_{\mathbf{t}}^{r\mathbf{t}}(\mathbf{x}) - L_{\mathbf{t}}^{r\mathbf{t}}(\mathbf{x}))dr$$

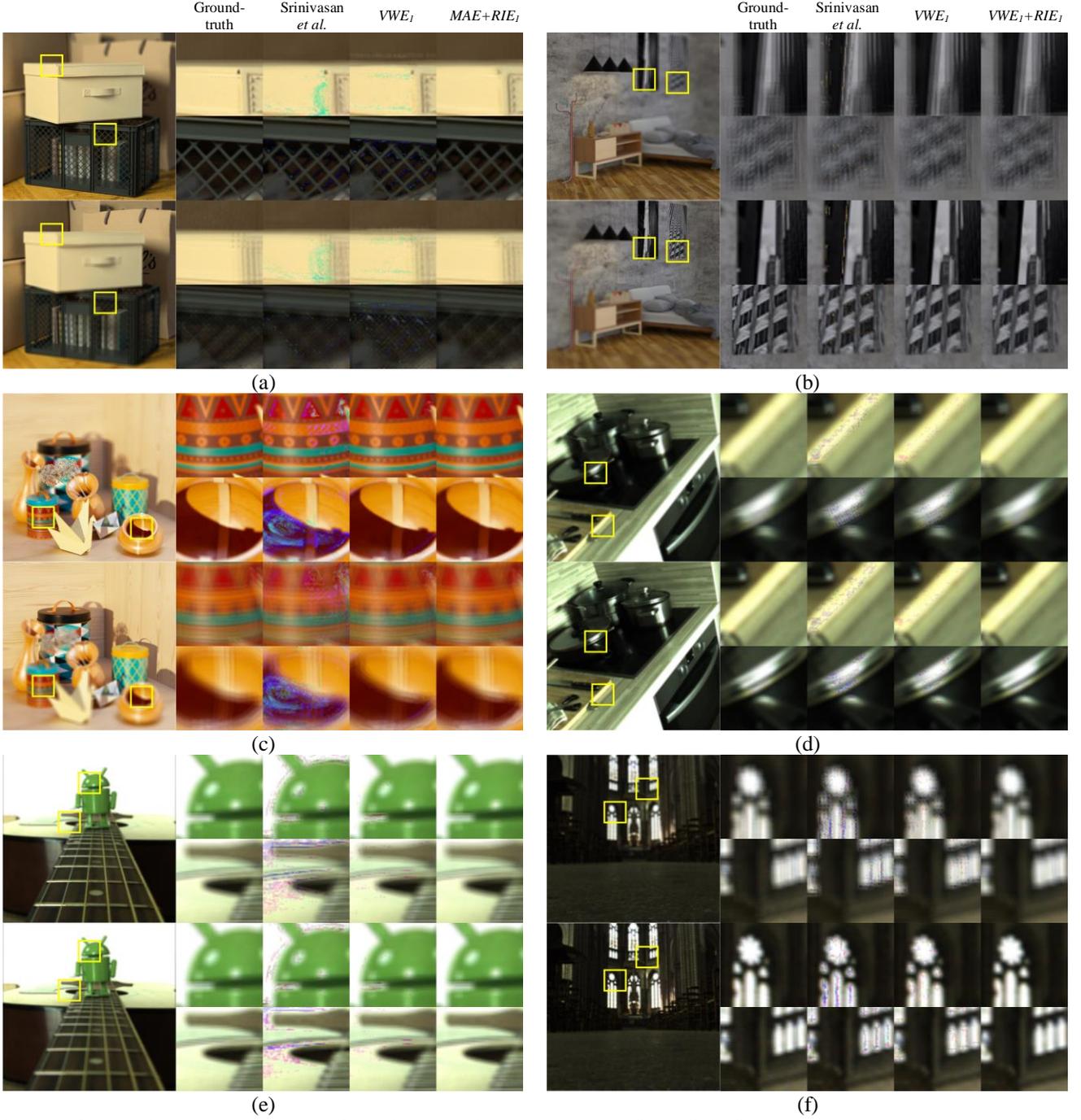

Fig. 7. Refocused images and close-ups from the ground-truth and the predicted light fields.

Let $\mathcal{E}_s = \mathfrak{F}\{\hat{L}_s - L_s\}$ for simplicity. By Plancherel's formula and the translation property of Fourier transform, we have

$$\frac{1}{2D(2N+1)^4}\int_{-D}^{D}\sum_{s,t}\sum_{x}(\hat{L}_s^{rs}(\mathbf{x})-L_s^{rs}(\mathbf{x}))(\hat{L}_t^{rt}(\mathbf{x})-L_t^{rt}(\mathbf{x}))dr$$

$$=\frac{1}{2D(2N+1)^4}\int_{-D}^{D}\sum_{s,t}\sum_{\omega}(\mathcal{E}_s(\omega)e^{jr\omega^T s})(\mathcal{E}_t(\omega)e^{jr\omega^T t})dr$$

The final step is to interchange the summation and the integration again,

$$\frac{1}{2D(2N+1)^4}\int_{-D}^{D}\sum_{s,t}\sum_{\omega}(\mathcal{E}_s(\omega)e^{jr\omega^T s})(\mathcal{E}_t(\omega)e^{jr\omega^T t})dr$$

$$=\frac{1}{(2N+1)^4}\sum_{s,t}\sum_{\omega}\mathcal{E}_s(\omega)\mathcal{E}_t(\omega)\frac{1}{2D}\int_{-D}^{D}e^{jr\omega^T(\mathbf{s}+\mathbf{t})}dr$$

$$=\frac{1}{(2N+1)^4}\sum_{s,t}\sum_{\omega}\mathcal{E}_s(\omega)\mathcal{E}_t(\omega)\operatorname{sinc}(D\omega^T(\mathbf{s}+\mathbf{t})).$$

For CRIE, we only need to replace $D$ with infinity and add $g(r)$ to the equation. Therefore, we have

$$CRIE_2(G_\theta(S), L) =$$
$$(\frac{1}{2N+1})^4 \sum_{s,t} \sum_{\omega} \mathcal{E}_s(\omega)\mathcal{E}_t(\omega) \frac{1}{2D} \int_{-\infty}^{\infty} e^{-r^2} e^{jr\omega^T(s+t)} dr)$$
$$= \frac{1}{(2N+1)^4} \sum_{s,t} \sum_{\omega} \mathcal{E}_s(\omega)\mathcal{E}_t(\omega) \frac{\sqrt{\pi}}{2D} e^{-0.25(\omega^T(s+t))^2}.$$

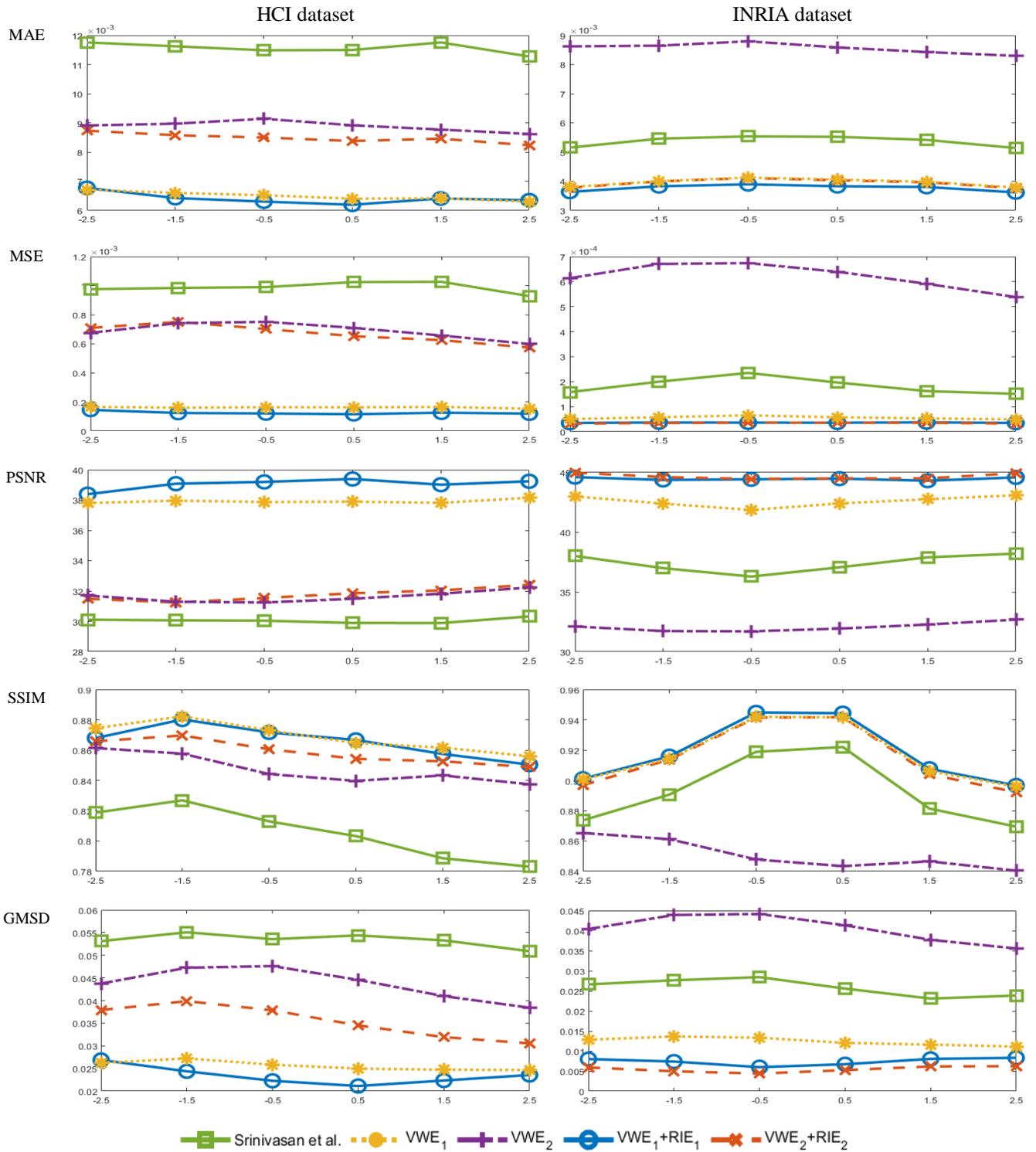

Fig. 8. MAE, MSE, PSNR, SSIM and GMSD scores of refocused images generated from the light fields predicted by five neural networks with $r$ ranging from $-2.5$ to $2.5$.

This completes the proof. □

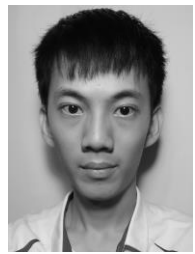

**Chang-Le Liu** is currently working toward the B.S. degree in electrical engineering in National Taiwan University, Taipei, Taiwan, from 2016.

He was a summer intern as a research assistant with the Research Center for Information Technology Innovation, Academia Sinica, Taipei, Taiwan, and was involved in research in speech enhancement. His current research topic includes image and audio signal processing and light field photography and rendering.

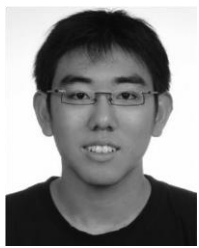

**Kuang-Tsu Shih** received the Ph.D. degree in communication engineering from National Taiwan University, Taipei, Taiwan, in 2017.

He is currently a Postdoctoral Researcher with the Graduate Institute of Communication Engineering, National Taiwan University. His current research interests include color image processing, color science, and computational photography.

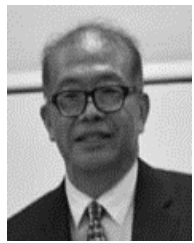

**Jiun-Woei Huang** is Research Fellow of Taiwan Instrument Research Institute, National Applied Laboratories, Taiwan, R.O.C., and adjunct associate professor, Institute of Applied Mechanics, National Taiwan University, since 2008. He received his BS and MS degrees in physics from the National Taiwan Normal University in 1977 and 1981, respectively, and his Ph.D. degree in Physics from Texas Christian University, USA, with Professor W.R.M. Graham in Molecular Spectroscopy, in May 1990. He worked as Post Doc of Lidar Laboratory of Professor C. Y. She in Colorado State University from May, 1990 to Aug. 1991, then back to Taiwan. From late 1991 to Jan. 1992. He was a Post-Doc Researcher in the Institute of Chemistry of Academia Sinica, R.O.C., studied for Electron Paramagnetic Resonance in minerals. Then he had been in CSIST from 1992 to 2014 in the field of optical design, mechanical structure, fiber gyro, and infrared optics. He is a member of SPIE, OSA, and life member of Sigma Xi.

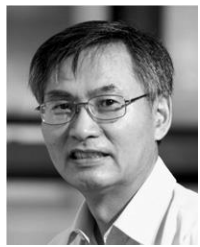

**Homer H. Chen** (S'83–M'86–SM'01–F'03) received the Ph.D. degree in electrical and computer engineering from the University of Illinois at Urbana–Champaign. His professional career has spanned industry and academia. Since August 2003, he has been with the College of Electrical Engineering and Computer Science, National Taiwan University, where he is a Distinguished Professor. Prior to that, he held various research and development management and engineering positions at U.S. companies over a period of 17 years, including AT&T Bell Labs, Rockwell Science Center, iVast, and Digital Island. He was a U.S. delegate for ISO and ITU standards committees and contributed to the development of many new interactive multimedia technologies that are now a part of the MPEG-4 and JPEG-2000 standards. His professional interests lie in the broad area of multimedia signal processing and communications.

Dr. Chen serves on the IEEE Signal Processing Society Awards Board and the Senior Editorial Board of the IEEE Journal on Selected Topics in Signal Processing. He was a Distinguished Lecturer of the IEEE Circuits and Systems Society from 2012 to 2013. He served on the IEEE Signal Processing Society Fourier Award Committee and the Fellow Reference Committee from 2015 to 2017. He was a General Chair of the 2019 IEEE International Conference on Image Processing. He was an Associate Editor of the IEEE Transactions on Circuits and Systems for Video Technology from 2004 to 2010, the IEEE Transactions on Image Processing from 1992 to 1994, and Pattern Recognition from 1989 to 1999. He served as a Guest Editor for the IEEE Transactions on Circuits and Systems for Video Technology in 1999, the IEEE Transactions on Multimedia in 2011, the IEEE Journal of Selected Topics in Signal Processing in 2014, and Multimedia Tools and Applications (Springer) in 2015.